\documentclass[12pt]{amsart}
\usepackage[utf8]{inputenc}
\usepackage{graphicx}
\usepackage{amssymb,hyperref}
\usepackage{color}

\title{The graphs behind Reuleaux polyhedra}

\author[Montejano]{Luis Montejano}
\author[Pauli]{Eric Pauli}
\author[Raggi]{Miguel Raggi}
\author[Roldán-Pensado]{Edgardo Roldán-Pensado}

\thanks{This reaserch was supported by PAPIIT projects IA102118, IN112614 and IN116919, and CONACyT project 166306}

\address{Instituto de Matemáticas, UNAM campus Juriquilla}
\email[L. Montejano]{luis@im.unam.mx}
\email[E. Pauli]{eriicpc@gmail.com}
\address{Escuela Nacional de Estudios Superiores, UNAM Campus Morelia}
\email[M. Raggi]{mraggi@gmail.com}
\address{Centro de Ciencias Matemáticas, UNAM Campus Morelia}
\email[E. Roldán-Pensado]{e.roldan@im.unam.mx}

\newtheorem{theorem}{Theorem}

\newtheorem{lemma}[theorem]{Lemma}

\newtheorem{conjecture}[theorem]{Conjecture}
\theoremstyle{definition}

\theoremstyle{remark}

\newcommand{\R}{\mathbb R}
\newcommand{\strongly}{strongly }

\DeclareMathOperator{\dist}{dist}

\begin{document}

\begin{abstract}
This work is about graphs arising from Reuleaux polyhedra. Such graphs must necessarily be planar, $3$-connected and \strongly self-dual. We study the question of when these conditions are sufficient.

If $G$ is any such a graph with isomorphism $\tau : G \to G^*$ (where $G^*$ is the unique dual graph), a \emph{metric mapping} is a map $\eta : V(G) \to \R^3$ such that the diameter of $\eta(G)$ is $1$ and for every pair of vertices $(u,v)$ such that $u\in \tau(v)$ we have $\dist(\eta(u),\eta(v)) = 1$. If $\eta$ is injective, it is called a \emph{metric embedding}. Note that a metric embedding gives rise to a Reuleaux Polyhedra.

Our contributions are twofold: Firstly, we prove that any planar, $3$-connected, \strongly self-dual graph has a metric mapping by proving that the chromatic number of the \emph{diameter graph} (whose vertices are $V(G)$ and whose edges are pairs $(u,v)$ such that $u\in \tau(v)$) is at most $4$, which means there exists a metric mapping to the tetrahedron. Furthermore, we use the Lovász neighborhood-complex theorem in algebraic topology to prove that the chromatic number of the diameter graph is exactly $4$.

Secondly, we develop algorithms that allow us to obtain \emph{every} such graph with up to $14$ vertices. Furthermore, we numerically construct metric embeddings for every such graph. From the theorem and this computational evidence we conjecture that \emph{every} such graph is realizable as a Reuleaux polyhedron in $\R^3$.

In previous work the first and last authors described a method to construct a constant-width body from a Reuleaux polyhedron. So in essence, we also construct hundreds of new examples of constant-width bodies.

This is related to a problem of Vázsonyi, and also to a problem of Blaschke-Lebesgue.
\end{abstract}

\maketitle

\section{Introduction} \label{sec:intro}

A \emph{Reuleaux polygon} $K\subset\R^2$ is the intersection of finitely many circles of radius $r$ in such a way that the \emph{vertices} of $K$ (\textit{i.e.} the non-smooth points on the boundary of $K$) are precisely the centers of the circles defining $K$. It is a simple exercise to see that for any Reuleaux polygon, $n$ must be odd and, conversely, that for every odd $n\ge 3$ there exist Reuleaux polygons with $n$ vertices.

A ball polyhedron $\Omega$ is the intersection of finitely many (at least three) $3$-dimensional balls of radius $1$ in $\R^3$. Ball polyhedra are natural objects used for studying several important problems in discrete geometry, (see \cite{KMP2010,BLNP2007}). The geometry of the boundary of a ball polyhedron is of particular interest. Indeed, one can represent the boundary of a ball polyhedron $\partial\Omega$ as the union of vertices, edges and faces defined in a natural way, giving rise to a graph $G_\Omega$ embedded in $\partial\Omega$ (see Section \ref{sec:poly}). The faces are the closure of the connected components of $\partial\Omega\setminus G_\Omega$, where each face is a closed subset of a sphere of radius $1$. Furthermore, each edge of $G_\Omega$ is an arc of a circle with radius smaller than $1$. This structure is not necessarily a lattice, unless the intersection of any two faces is either empty, a vertex or an edge. If this is the case, $\Omega$ is called a \emph{standard} ball polyhedron, and its corresponding graph $G_\Omega$ is a simple $3$-connected planar graph.

\begin{figure}
\begin{center}
\includegraphics[width=0.33\textwidth]{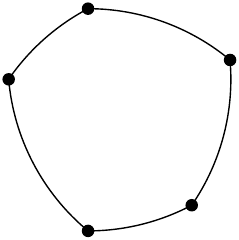}\qquad
\includegraphics[width=0.33\textwidth]{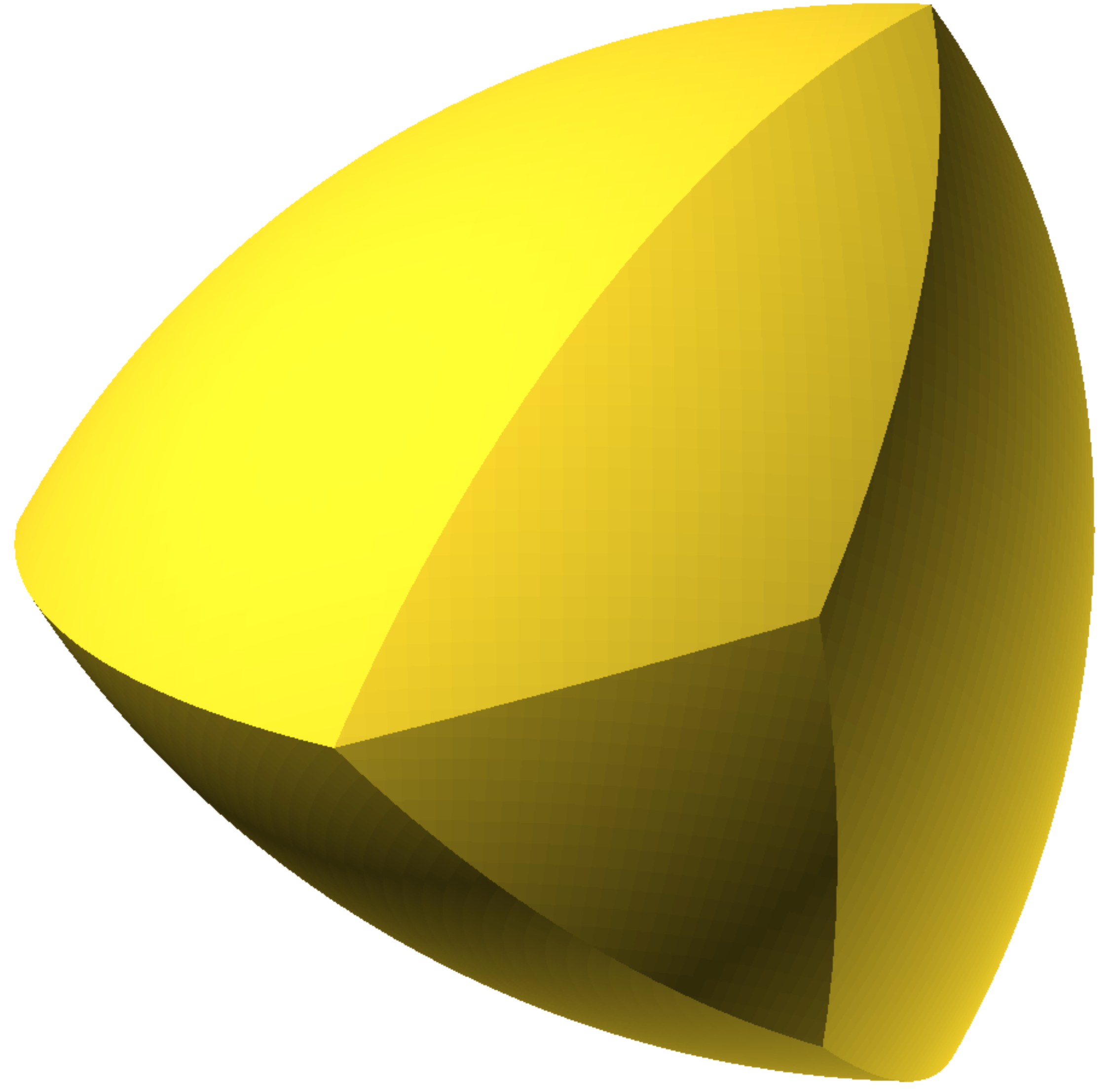}
\caption{A Reuleaux polyhedron and a Reuleaux polyhedron.}\label{fig:reuleaux}
\end{center}
\end{figure}

Let $\Omega$ be a standard ball polyhedron defined by a set $X$. In other words,
$$\Omega=\bigcap_{x\in X}B(x,1)$$
is the intersection of a collection of $3$-dimensional balls of diameter $1$ centered around the points of a finite set $X\subset\R^3$. There is a certain type of standard ball polyhedra which is of particular interest. We say that $\Omega$ is a \emph{Reuleaux polyhedron} if $\Omega$ has the property that the corner points of $\partial\Omega$, and hence the vertices of the graph $G_\Omega$, are exactly the points of $X$ (see Figure~\ref{fig:reuleaux}).

Reuleaux polyhedra are of importance in convex and discrete geometry for many reasons, but in particular for the following two.

We wish to characterize the finite sets $X \subset \R^3$ for which the diameter is attained a maximal number of times as a segment of length $1$ with both endpoints in $X$. Vázsonyi conjectured that a finite set $X$ of size $m$ and diameter $1$ has at most $2m-2$ diameters. We say that $X\subset\R^3$ is an \emph{extremal set} if $X$ has diameter $1$ and has exactly $2m-2$ diameters. Furthermore, $X\subset\R^3$ is a Vázsonyi set if in addition every point of $X$ is in at least $3$ diameters. It is not difficult to see that every extremal set can easily be obtained from a Vázsonyi set. Grünbaum, Heppes, and Straszewicz proved independently, using ball polyhedra, that a Vázsonyi set is precisely a Reuleaux set. In other words, given a Vázsonyi set $X$, the intersection of balls of radius $1$ centered at $X$ is a Reuleaux polyhedron and, furthermore, the set of corner points on the boundary of a Reuleaux polyhedron is an extremal Vázsonyi set (see \cite{CG1983}, \cite{MMO2019}).

The second reason for our interest in these objects is that there is a procedure which can be used to transform a Reuleaux polyhedron into a body of constant width $1$. This procedure is described in \cite{MMO2019} and follows the spirit in which Meissner transformed the Reuleaux tetrahedron into a body of constant width. The bodies of constant width obtained from Reuleaux polyhedra are called \emph{Meissner polyhedra}. The boundary of a Meissner  polyhedron is made out of spherical caps of radius $1$ and surfaces of revolution over arcs of circles with radius $1$. In particular, this implies that a Meissner body is a body of constant width $1$ with the property that the smooth components of its boundary have their smaller principal curvature constant.

Also worth mentioning here is the Blaschke-Lebesgue problem, which consists of minimizing the volume in the class of convex bodies of constant width $1$. Anciaux and Guilfoly \cite{AG2011} proved that the minimizer of the Blaschke-Lebesgue problem has this curvature property and therefore Meissner bodies are optimal candidates to be solutions of this problem. 

\vspace{1em}

Let $G=(V,E)$ be a planar $3$-connected self-dual graph. Suppose that the self-duality isomorphism $\tau: G \to G^*$ is \strongly involutive. That is, it satisfies the following two properties:
\begin{enumerate}
\item For every vertex $u$, $u\notin \tau(u)$.
\item For every pair of vertices $u,v$, we have that $u\in\tau(v) \iff v\in\tau(u)$.
\end{enumerate}

There are many examples of \strongly involutive self-dual graphs. For example, wheels with an odd number of sides. For another example, see Figure \ref{fig:eight}.

A \emph{metric mapping} of a \strongly involutive self-dual graph $G$ is a map $\eta:V\to \R^3$ such that the diameter of $\eta(V)$ is $1$, and for every $v\in V$ and $u\in\tau(v)$ we have $\dist(\eta(u),\eta(v))=1$. If, in addition, $\eta:V\to \R^3$ is injective, then it is called a \emph{metric embedding}.

\begin{figure}
\begin{center}
\includegraphics[width=0.5\textwidth]{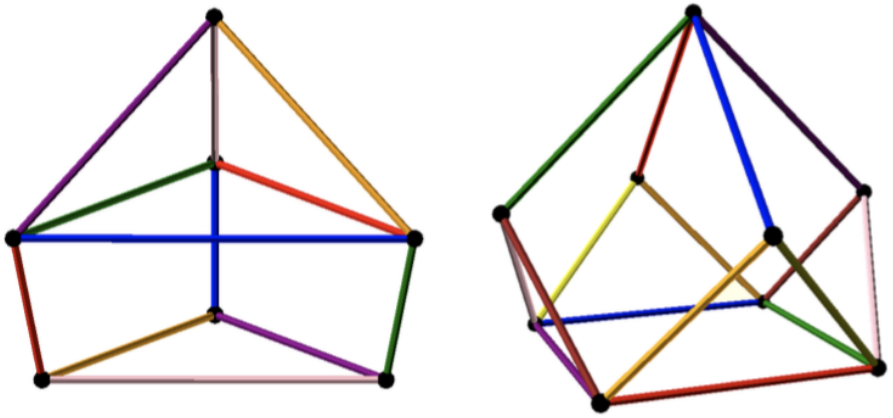}
\caption{Metric embeddings of two \strongly involutive self-dual graphs.}\label{fig:metric}
\end{center}
\end{figure}

One can construct examples of metric embeddings for wheels with an odd number of sides, the graphs of Figure \ref{fig:metric} and some other families of \strongly involutive self-dual graphs (see \cite{MR2017}). In particular, the spherically self-dual polyhedra constructed by Lovász in \cite{Lov1983a}, which is related to the Erdős-Graham problem, are also examples of metric embeddings.

Given a metric embedding $\eta$, let $\Omega$ be the ball polyhedron defined by the set $\eta(V)$. Then $\Omega$ is a Reuleaux polyhedron.

Moreover, the face structure of the boundary complex of $\partial\Omega$ induced by the graph $G_\Omega$ is isomorphic to the face structure of the embedding of $G$ as a map in $\mathbb{S}^2$. Conversely, given a Reuleaux polyhedron $\Omega=\bigcap_{x\in X}B(x,1)$, the graph $G_\Omega$ is a \strongly involutive self-dual planar graph and $X$ is the image of a metric embedding of $G_\Omega$. See \cite{MR2017} and Chapter 6 of \cite{MMO2019}.

Consequently, in order to construct and classify Reuleaux polyhedra it is natural to ask if every \strongly involutive self-dual graph admits a metric embedding.

Our belief is that the answer is affirmative, as stated in Conjecture \ref{conj:reuleaux}. Since we have not been able to prove this conjecture, we give a weaker version in the following theorem, which we prove in Section \ref{sec:r-c}.

\begin{theorem}\label{thm:metric}
	Given a \strongly involutive self-dual graph $G$, there exists a (not necessarily injective) metric mapping of $G$.
\end{theorem}

Finally, in Section \ref{sec:comp}, we describe a computer program that finds all \strongly involutive self-dual graphs with up to $14$ vertices. The number of \strongly involutive self-dual graphs with a given number of vertices is detailed in Table~\ref{table}.

\begin{table}
	\begin{tabular}{|l|c|c|c|c|c|c|c|c|c|c|c|}
		\hline
		\textbf{\# of vertices} & 4 & 5 & 6 & 7 & 8 & 9 & 10 & 11 & 12 & 13 & 14 \\
		\hline
		\textbf{\# of graphs} & 1 & 0 & 1 & 1 & 2 & 4 & 11 & 24 & 72 & 212 & 674\\
		\hline
	\end{tabular}
	\vspace{6pt}
	\caption{Number of \strongly involutive self-dual graphs.}\label{table}
\end{table}

Once we have these graphs, we succeed in numerically constructing Reuleaux polyhedra in $\R^3$ from these graphs. We thus provide computational evidence supporting the following conjecture.
\begin{conjecture}\label{conj:reuleaux}
	For every planar $3$-connected \strongly involutive self-dual graph $G$ there is a Reuleaux polyhedron $\Omega$ such that $G_\Omega$ is isomorphic to $G$.
\end{conjecture}

\section{Ball polyhedra and Reuleaux polyhedra}\label{sec:poly}

Let us define and point out some properties of \emph{ball polyhedra}. To read more about these, we recommend \cite{BLNP2007}, \cite{BN2006} and especially \cite{KMP2010}. 

For $x\in\R^3$ and $r>0$ we denote by $B(x,r)$ and $S(x,r)$ the ball and sphere with center $x$ and radius $r$, respectively. A \emph{ball polyhedra} $\Omega$ is simply the intersection of finitely many balls of radius $1$, so we may write $K=\bigcap_{x\in X} B(x,1)$ for some finite set $X\subset\R^3$.

Let us assume that $\Omega$ has non-empty interior and that $X$ is a minimal set that describes $\Omega$, \textit{i.e.} it has no redundant points. Then a point $x$ in the boundary $\partial \Omega$ of $\Omega$ can be of three different types:
\begin{itemize}
 \item We say that $x$ belongs to a face if it is a smooth point of $\partial \Omega$ or, equivalently, if $S(x,1)\cap X$ has exactly $1$ point.
 \item We say that $x$ belongs to an edge if $S(x,1)\cap X$ has at least $2$ points and is contained in some great circle of $S(x,1)$.
 \item We say that $x$ is a vertex if $S(x,1)\cap X$ is not contained in any great circle of $S(x,1)$.
\end{itemize}

In this way we define the faces, edges and vertices of $\Omega$ as with a cell-complex. The vertices and the edges are together the singular points of $\partial\Omega$, where the vertices are precisely the corners or singular vertex points of $\partial\Omega$. Together, the vertices and edges define a graph $G_\Omega$ embedded in $\partial\Omega$. 

All other points in $\partial\Omega\setminus G_\Omega$ are smooth points of $\partial\Omega$. The faces are hence the closure of the connected components of $\partial\Omega\setminus G_\Omega$, where each face is a closed subset of a sphere of radius $r$ and, furthermore, each edge of $G_\Omega$ is an arc of a circle of radius smaller than $r$. 

Note that this structure does not necessarily form a lattice, unless the intersection of any two faces is either empty, a vertex or an edge. If this is the case, $\Omega$ is called a \emph{standard ball polyhedron} and its corresponding graph of singularities $G_\Omega$ is a simple $3$-connected planar graph.

A \emph{Reuleaux polyhedron} is a standard ball polyhedron in which the set of vertices coincides with the set of centers (see \cite{MR2017}). For example, see Figure~\ref{fig:reuleaux}.

\section{Involutive self-dual graphs} \label{sec:graphs}
Given a planar graph $G$, any regular embedding of the topological realization of $G$ into the sphere $\mathbb{S}^2$ partitions the sphere into regions called the \emph{faces} $F$ of the embedding. Call this embedding $M=(V,E,F)$ a \emph{map}. We do not make a significant distinction between the geometric and combinatorial objects.

Given a map $M$, form the \emph{dual map} $M^*$ as usual, by placing a vertex $f^*$ in the center of each face $f$, and for each edge $e$ of $M$ draw a dual edge $e^*$ connecting the vertices $f_1^*$ and $f_2^*$ by crossing $e$ transversely, if the two faces that contain $e$ are $f_1$ and $f_2$.

Note that if $G$ is $3$-connected there is a unique embedding in the sphere (up to isomorphism), so the dual is fully determined by the graph alone.

A map $M$ is \emph{self-dual} if there is a map isomorphism
    $$\tau:M \to M^*.$$

Therefore, $\tau$ associates to each vertex of $M$ a vertex of $M^*$, which corresponds to a face of $M$. Since $\tau$ is fully determined by $\tau|_V : V \to V(M^*)=F(M)$, we may sometimes abuse notation and also refer to the restriction as $\tau$. Moreover, if $v$ is a vertex, we think of $\tau(v)$ as the set of vertices corresponding to a face of $M$.

Also note that a self-dual map with $n$ vertices must have $2n-2$ edges, which follows from Euler's formula. For more about self-dual graphs see \cite{SS1996}.

We define the \emph{map of squares} $M^\square$ (and the corresponding \emph{graph of squares} $G^\square$) as the ``union'' of $M$ and $M^*$: In order to have a map, we must place a vertex (of degree 4) at the intersection edges of $M$ and $M^*$. More formally: vertices of $M^\square$ are either vertices of $M$, vertices of $M^*$, or points at the intersection of an edge of $M$ with its dual edge in $M^*$. Edges of $M^\square$ are half edges of $M$ or $M^*$, split by their intersection.

It a known result that interestingly the embedding can be chosen in such a way that every automorphism of the graph of $M$ comes from an isometry of $\mathbb{S}^2$. Furthermore, the embeddings of $M$ and $M^*$ can be chosen in such a way that any automorphism of $G^\square$ comes from an isometry of $\mathbb{S}^2$.

If $M$ is self-dual with isomorphism $\tau$, note that $\tau$ can be thought of as an automorphism of $M^\square$ and therefore can be viewed as an isometry of $\mathbb{S}^2$.

We say that a self-dual map $M$ is \strongly involutive if isomorphism $\tau:M\to M^*$ satisfies the following two properties:
\begin{enumerate}
\item $v\notin \tau(v)$ for every $v\in V$,
\item $u\in \tau(v) \iff v\in \tau(u)$.
\end{enumerate}
The second property implies that $\tau^2=Id$ when viewed as an automorphism of $M^\square$.

Interestingly, 1) and 2) together imply that the embeddings of $M$ and $M^*$ can be chosen in such a way that the automorphism of $\tau$ is the antipode of $\mathbb{S}^2$.

\section{Metric embeddings} \label{sec:metric}

In what follows, $M$ is a \strongly involutive self-dual map with isomorphism $\tau$ with underlying graph $G$.

\begin{figure}
\begin{center}
\includegraphics[scale=0.6]{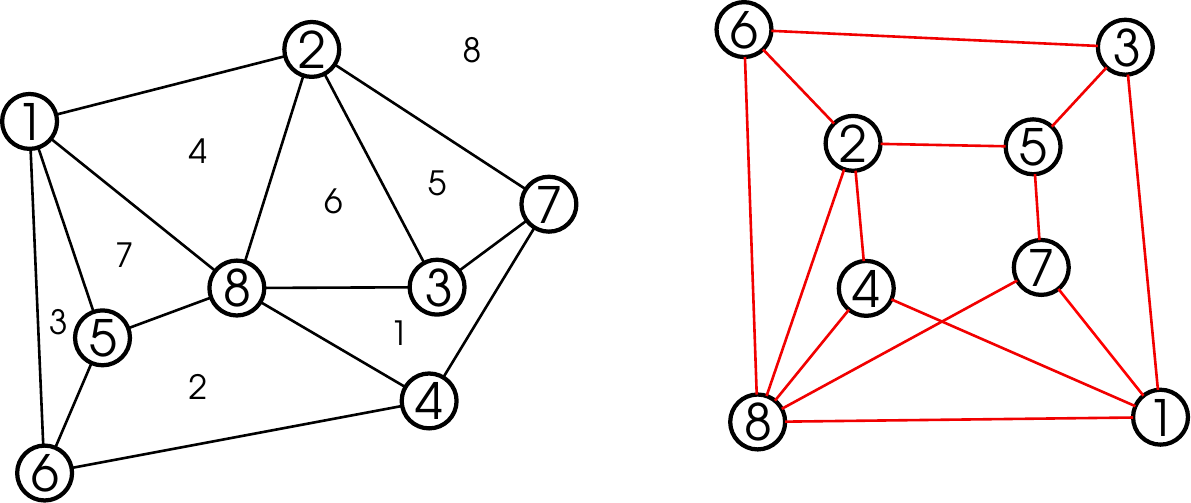}
\caption{A \strongly involutive self-dual graph (left) and its diameter graph (right). The numbers on the faces represent the isomorphism $\tau$.}\label{fig:eight}
\end{center}
\end{figure}

Define the \emph{diameter graph} $D(M)$ as a graph whose vertices are the vertices of $M$, but whose edges are pairs of vertices $\{u,v\}$ such that $v\in \tau(u)$. See Figure~\ref{fig:eight}.

A \emph{metric mapping} of $M$ is a function $\eta:V\to \R^3$ such that the edges of $D(M)$ have length 1 and the diameter of $\eta(V)$ is $1$. In other words, the distance from a vertex to every vertex of its opposite face is exactly $1$. Note that we do not require $\eta$ to be injective in this definition. If the mapping is injective, we call it a \emph{metric embedding}.

Given $\eta$ a metric embedding of a \strongly involutive self-dual graph $M$, let 
$$\Omega=\bigcap_{v\in V}B(\eta(v),1).$$

Note that the points of $\eta(V)$ exactly coincide with the singular points of $\partial\Omega$. Moreover, the face structure of the boundary complex of $\partial \Omega$ is isomorphic to the face structure of $M$ (see \cite{KMP2010}, \cite{MR2017} and \cite[Chapter 6]{MMO2019}).

\section{The remove-contract operation and the chromatic number of $D(M)$} \label{sec:r-c}

It is natural then to ask if every \strongly involutive self-dual graph admits a metric embedding. Our belief is that the answer is affirmative, as stated in Conjecture \ref{conj:reuleaux}. Since we have not been able to prove this conjecture, we give a weaker version in Theorem~\ref{thm:metric}. In fact, we prove the following theorem, and Theorem~\ref{thm:metric} follows immediately.

\begin{theorem}\label{thm:chromatic}
	The diameter graph $D(M)$ has chromatic number exactly $4$. Therefore, there exist a metric mapping $\eta:V(M) \to X$ where $X\subset \R^3$ are the vertices of a regular tetrahedron of side 1. That is, for every vertex $v\in V(G)$ and every vertex $u\in\tau(v)$ we have that $\dist(\eta(u),\eta(v)) = 1$.
\end{theorem}

This theorem in turn implies Theorem~\ref{thm:metric}, that a metric mapping of $M$ always exists.

Additionally, together with some theorems about rigid graphs, this result provides strong evidence that a metric embedding does indeed always exist: It follows from the work done in \cite{AR1978} that, since $D(M)$ has $2n-2 < 3n-6$ edges, the tetrahedral metric mapping of $D(M)$ mentioned in the theorem is not rigid, and in fact has several degrees of freedom. However, this does not guarantee that there is actually a metric embedding of $M$.

In order to prove Theorem~\ref{thm:chromatic}, let us first define the remove-contract operation as follows:

Suppose $M$ is a \strongly involutive self-dual map with underlying graph $G$ and suppose $ab$ is an edge of $M$. Then by definition there are two faces $\tau(a)$ and $\tau(b)$ and an edge $xy$ in the intersection of $\tau(a)$ and $\tau(b)$ such that $xy^*=\tau(ab)$. Note that the two edges $ab$ and $xy$ are disjoint.

\begin{figure}
\begin{center}
\includegraphics[scale=0.6]{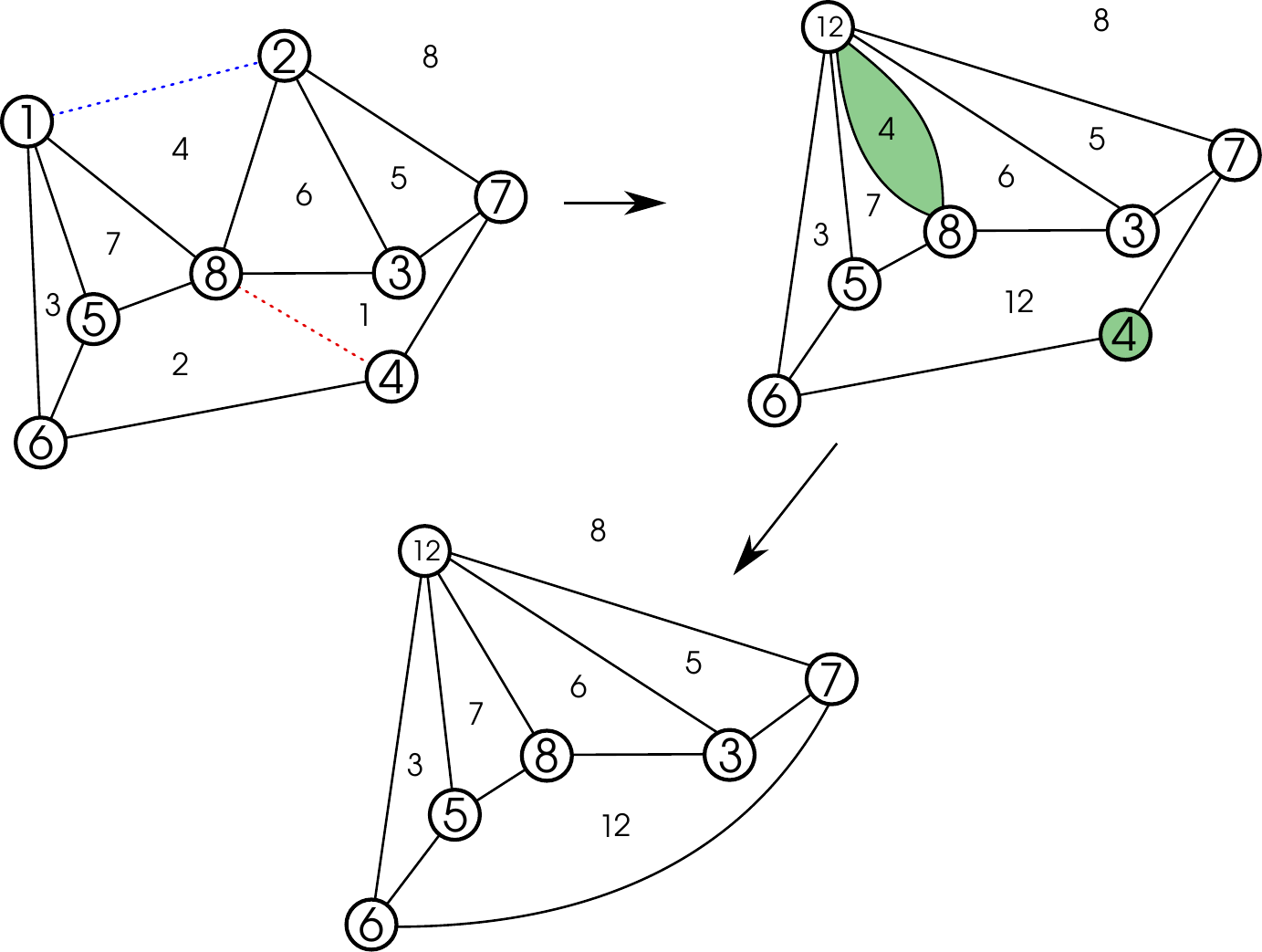}
\caption{The remove-contract operation. Remove edge 4-8 and contract edge 1-2. Vertex 4 has degree $2$ (and face 4 has two edges), so remove them as shown. }\label{fig:r-c}
\end{center}
\end{figure}

Consider the following self-dual map: contract edge $ab$ and at the same time delete the edge $xy$, thereby creating a bigger face with the union of $\tau(a)$ and $\tau(b)$. Then erase any vertex of degree two and contract any face with two edges. See Figure \ref{fig:r-c} for an illustration.

More formally, the remove-contract operation in $M$ with edge $ab$ gives rise to a map $M^\diamond= M^\diamond_{ab}$, where the underlying graph $G^\diamond$ is obtained from graph $G$ simply by deleting edge $xy$ and contracting the edge $ab$. Then define $\tau^\diamond (v)=\tau(v)$ for every $v\notin\{a,b\}$ and let $\tau^\diamond (a=b)$ be the face obtained by the union of $\tau(a)$ and $\tau(v)$ when edge $xy$ is deleted.

Clearly $M^\diamond$ is a self-dual map satisfying property 2). Furthermore if edge $ab$ is not an edge of $D(M)$, then it is easy to verify that $M^\diamond$ is again a \strongly involutive self-dual map. The only property that might be violated is the $3$-connectedness, but any such violation is easily remedied by simply removing vertices of degree $2$ and contracting faces with exactly two edges. Clearly, $\tau^\diamond$ is still \strongly involutive self-dual when restricted to the new map.

The idea now is to repeatedly apply the operation to $M$ by selecting edges of $M$ that are not edges of $D(M)$.

In light of this, we propose the following lemma.
\begin{lemma}
If every edge of $M$ is also an edge of $D(M)$, then $M=K_4$.
\end{lemma}

In other words, there exists an edge of $M$ that is adequate to perform the remove-contract operation, unless $M=K_4$.

\begin{proof}
Note that for every $v\in V(M)$, the degree of $v$ in $M$ is exactly the degree of $v$ in the diameter graph $D(M)$. Therefore, $M$ and $D(M)$ have the same number of edges. Thus, if every edge of $M$ is an edge of $D(M)$, then the edges of $M$ and $D(M)$ exactly coincide.

Assume then that the edges of $M$ and $D(M)$ coincide. Let $ab$ be an edge of $M$ (and so also of $D(M)$) and let $xy = \tau(ab)^*$ an edge of $M$ (and so also in $D(M)$). Consider faces $\tau(a)$ and $\tau(b)$. By the properties of $\tau$, $a\notin \tau(a)$ and $b \notin \tau(b)$. But since $ab\in D(M)$, $b\in \tau(a)$ and $a\in\tau(b)$. This means $a,b,x,y$ form a tetrahedron and are therefore the only vertices in $M$.
\end{proof}

We are now ready to prove Theorem~\ref{thm:chromatic} that the chromatic number of $D(M)$ is 4.

\begin{proof}
To prove that $4$ colors suffice, color each vertex according to the vertex of the tetrahedron $K_4$ to which they ended up identified after repeatedly applying the remove-contract operation. Since no edge of $D(M)$ was selected for the operation, two vertices that were identified could not have an edge of $D(M)$ between them.

We shall prove next that the chromatic number of $D(M)$ cannot be less than $4$. For that purpose we use a topological result of Lovász, who gives the following lower bound on the chromatic number of a graph: Let $G=(V,E)$ be a finite graph. Define its neighborhood simplicial complex $N(G)$ as the simplicial complex with vertices $V$, and where a subset $A\subset V$ forms a simplex of $N(G)$ if and only if the the vertices of $A$ have a common neighbor. In our case, if $D(M)$ is the diameter graph of a \strongly involutive self-dual graph $M$, the simplicial complex $N(D(M))$ is obtained by introducing a simplex for each face of $M$. Consequently, $N(D(M))$ has the homotopy type of the sphere $\mathbb{S}^2$. In \cite{Lov1978}, it was proved that if for a graph $G$ its neighborhood simplicial complex $N(G)$ is $k-connected$, then $\chi(G)\geq k+3$. Since $N(D(M))$ has the homotopy type of the sphere $\mathbb{S}^2$, then it is $1$-connected and therefore $\chi(N(D(M))\geq 4$ as we wished. A very similar argument was given previously in \cite[proof of Theorem 2]{Lov1983a}.
\end{proof}

\section{Finding \strongly involutive self-dual graphs and their metric embeddings} \label{sec:comp}

In this section we describe the computational algorithms used to construct Reuleaux polyhedra from all \strongly involutive self-dual graphs with up to $14$ vertices. This software was written mostly in \texttt{C++} and can be freely downloaded (and used) from
\begin{center}
 \url{https://github.com/mraggi/ReuleauxPolyhedra}.
\end{center}

Our pipeline includes 4 steps:
\begin{enumerate}
	\item Generate all 3-connected planar graphs with the appropriate number of edges.
	\item Select only \strongly involutive self-dual graphs.
	\item Embed the diameter graphs in $\R^3$.
	\item Create a Meissner or Reuleaux polyhedron from the embedding, for visualization.
\end{enumerate}

\subsection{Generating 3-connected planar graphs}

The process starts by using \texttt{plantri} \cite{BM2007}, Brinkmann and McKay's wonderful software. We ask \texttt{plantri} to generate all $3$-connected planar graphs with $n$ vertices (with $n \leq 14$) and exactly $2n-2$ edges. This is by far the slowest step in the pipeline and therefore the bottleneck. This means that further optimizations in the following steps are not necessary.

\subsection{Constructing involutive isomorphisms}

Once we have a list of all planar $3$-connected graphs with the appropriate number of edges, we wish to find an bijection $\tau$ from the set of vertices $V(G)$ to the set of faces $F(G)$ that satisfies the properties of \strongly involutive self-dual graphs detailed in Section~\ref{sec:graphs}.

We accomplish this by posing the problem as a CSP (constraint satisfaction problem) and using the arc-consistency algorithm \cite{Mac1977} in order to reduce the search space, followed by standard search. This works remarkably well in practice, and in fact takes virtually no time to refine the list generated in the previous step and find all \strongly involutive self-dual graphs with up to $14$ vertices. This is remarkable considering there are 23,556 planar $3$-connected graphs with $14$ vertices (and only $674$ of them are \strongly  involutive self-dual).

This is how the arc consistency algorithm works in this context: Let $G$ be a $3$-connected planar graph. For each vertex $v \in V(G)$, associate a set-like data structure of candidate faces $F_v$, denoting ``possible mappings''. At first, $F_v$ consists of all faces $f$ with the same number of edges as the vertex degree and for which $v \notin f$ (in order to satisfy the definition of involutive self-dual graph), but we shall reduce the search space by repeatedly discarding faces which could not possibly be mapped to $v$. We worry about making $\tau$ a strong involution later, and focus now on simply constructing an isomorphism to the dual.

Consider all edges $uv \in E(G)$ (in CSP terms, these correspond to arcs $\text{variable}\to\text{condition}$). Place them initially in a set-like data structure called \texttt{EdgesToProcess}. The arc-consistency part of the algorithm ends when this data structure is empty. Once an edge is removed from \texttt{EdgesToProcess} we say it is (temporarily) \emph{consistent}.

Repeatedly consider edges $uv$ from \texttt{EdgesToProcess} and make them consistent as follows: For each face $f \in F_u$, see if there exists a face $g\in F_v$ such that $fg\in E(G^*)$. This is in order for $\tau$ to stand a chance of being an isomorphism. If there is no such $g$, then remove $f$ from $F_u$, as we are certain $\tau(u) \neq f$.  Of course, we must now add all edges $xu \in E(G)$ to \texttt{EdgesToProcess}, as they might have stopped being consistent in the case $f$ was needed to make $xu$ consistent. If there is such a $g$, simply proceed to the next edge.

While we could further restrict the search space by performing a similar process for the property that $\tau$ must be strongly involutive, we found no further optimizations were useful for $n \leq 14$.

Once the above process finishes, we search all possible candidate $\tau$ by creating a tree of partial assignments and branching for each vertex $v$ with every member of $F_v$, and pruning the tree when a partial assignment leads to inconsistencies, either because the an isomorphism or the \strongly involutive properties are violated.

Given an isomorphism $\tau$ we can easily construct the diameter graph, which is helpful for the next steps.

\subsection{Finding an embedding}

Once we have a involutive self-dual graph $M$ and its diameter graph $D(M)$, we embed it in $\R^3$ by choosing an appropriate objective function, for which we find minima using differential evolution \cite{Sto1996,SP1997}.

Our software is able to numerically construct a good embedding for each involutive self-dual graph in a couple of seconds per graph. Bear in mind we want mappings with three properties: distance $1$ for edges in $D(M)$, non degeneracy, and diameter 1.

In other words, we wish for the length of every edge of the diameter graph to be as close as possible to $1$. Secondly, we also require that all the other pairs of vertices are somewhat separated so as to not construct degenerate or nearly degenerate examples. Finally, no pair must be at distance grater than $1$ so that the overall diameter does not exceed $1$.

Let $\eta:V(M)\to\R^3$ be a possible metric embedding of the diameter graph $D=D(M)$. For a pair of points $a,b \in \eta(V)$, define $h$ and $w$ as follows:
$$h(a,b) = \begin{cases}
	1 & \text{if }\dist(a,b) < \varepsilon, \\
	0 & \text{otherwise,} \\
\end{cases}
\text{ and }
w(a,b) = \begin{cases}
	1 & \text{if }\dist(a,b) > \alpha, \\
	0 & \text{otherwise,} \\
\end{cases}$$
where $\varepsilon=0.2$ and $\alpha=0.95$.

With the previous considerations in mind, if $\hat a = \eta(a)$, the objective function we attempted to minimize is
$$J_M(\eta) = \sum_{ab\in E(D)} (\dist(\hat a,\hat b)^2-1)^2 + K\sum_{ab\notin E(D)} h(\hat a,\hat b) + w(\hat a,\hat b)$$
where $K=10$. The values of $\varepsilon$, $\alpha$ and $K$ were chosen for practical reasons and seem to work well. Setting a higher value for $\varepsilon$ (\textit{e.g.} $0.3$), did not yield an embedding for every graph in our software.

We stopped the algorithm when the value of $J_M$ was less than $10^{-14}$. This means, in particular, that no pair of points are too close to each other and, moreover, that the length of edges of $D$ is almost 1, with an average error of about $0.0001$.

This experimental evidence is what led us to venture Conjecture~\ref{conj:reuleaux}.

\subsection{Visualization}

\begin{figure}
\begin{center}
\includegraphics[width=0.33\textwidth]{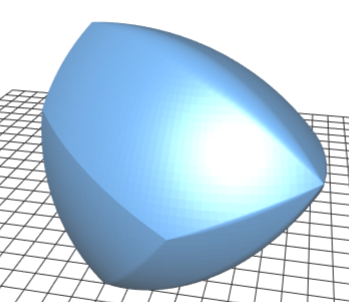}
\qquad
\includegraphics[width=0.33\textwidth]{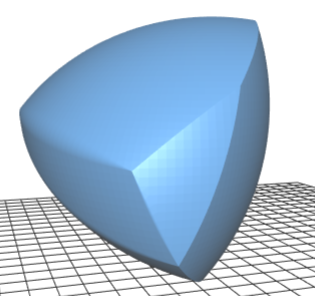}
\caption{A Meissner body (from two different angles).}\label{fig:meissner}
\end{center}
\end{figure}

Lastly, we include code that allows us to construct a 3D model of either a Reuleaux polyhedron or a Meissner polyhedron from the embeddings found. It is a small script written in \texttt{OpenSCAD} (\url{https://www.openscad.org/}). For the Reuleaux polyhedron it simply constructs the intersection of the corresponding spheres. To construct a Meissner polyhedron we follow the procedure described in \cite{MR2017}. For convenience, we include premade \texttt{STL} files of one Meissner polyhedron corresponding to each involutive self-dual graph with up to $11$ vertices. Figure~\ref{fig:meissner} has a rendering of the Meissner body associated with the example in Figure~\ref{fig:eight}. For others, see the github site.

\end{document}